\documentclass[12pt,epsfig]{article}
\usepackage{latexsym,amsmath,amssymb,epsfig,graphicx}

\newcommand{\be}{\begin{equation}}
\newcommand{\ee}{\end{equation}}
\newcommand{\bea}{\begin{eqnarray}}
\newcommand{\eea}{\end{eqnarray}}
\newcommand{\beast}{\begin{eqnarray*}}
\newcommand{\eeast}{\end{eqnarray*}}
\newcommand{\pkt}{\;\;.}
\newcommand{\kma}{\;\;,}
\newcommand{\tr}{{\rm tr}}

\newcommand{\calm}{{\cal M}}
\newcommand{\calv}{{\cal V}}
\newcommand{\calj}{{\cal J}}
\newcommand{\calg}{{\cal G}}
\newcommand{\cald}{{\cal D}}
\newcommand{\bfM}{{\bf M}}
\newcommand{\bfV}{{\bf V}}
\newcommand{\bfJ}{{\bf J}}
\newcommand{\bfG}{{\bf G}}
\newcommand{\bfI}{{\bf I}}
\newcommand{\bfx}{{\bf x}}
\begin{document}
\begin{titlepage}
\begin{flushright}
hep-th/yymmnn \\
March 2008
\end{flushright}
\vspace{20mm}
\begin{center}
{\Large \bf
One-loop corrections to the instanton transition
in the Abelian Higgs model: Gel'fand-Yaglom and Green's function methods }

\vspace{10mm}

{\large  J\"urgen Baacke\footnote{e-mail:~
juergen.baacke@tu-dortmund.de}} \\
\vspace{15mm}

\vspace{15mm}

{\large  Fachbereich Physik, Technische Universit\"at Dortmund \\
D - 44221 Dortmund, Germany
}
\vspace{15mm}

\newpage
\bf{Abstract}
\end{center}
The fluctuation determinant, the preexponential factor for the instanton
transition, has been computed several years ago
in the Abelian Higgs model, using a method based 
on integrating the Euclidean Green' function.
A more elegant method for computing functional determinants, 
using the Gel'fand-Yaglom theorem, has been 
applied recently to a variety of systems. This method runs into
difficulties if the background field has nontrivial topology, as 
is the case for the instanton in the Abelian Higgs model.
A shift in thre effective centrifugal barriers makes the s-wave
contribution infinite, an infinity that is compensated by the 
summation over the other partial waves.
This requires some modifications of the Gel'fand-Yaglom method
which are the main subject of this work.
We present  here both, the Green' s function and the Gel'fand-Yaglom
method and compare the numerical results in detail. 
\end{titlepage}

%*******************************************************Introduction

\section{Introduction}\label{intro}
\par

The computation of functional determinants for various background
field configurations has recently found renewed interest.
In particular, the elegant Gel'fand-Yaglom
\cite{Gelfand:1960,vanVleck:1928,Cameron:1945,Dashen:1974} approach,
denoted sometimes as ``the Coleman method''
(as presented in \cite{Coleman85}),
has been considered by various authors
\cite{Baacke:2003uw,Kirsten:2003py,
Kirsten:2004qv,Burnier:2005he,
Dunne:2005rt,Dunne:2006ct,Dunne:2007mt}, following earlier work
on vacuum decay and bubble nucleation
\cite{Kiselev:1975eq,Kiselev:1985er,Selivanov:1988jx,
Baacke:1993ne,Baacke:1995bw,Surig:1997ne}.
 Another method for computing the functional determinant, based on the
integration of the Euclidean Green's function,
 has been used in Refs.
\cite{Baacke:1994bk} and \cite{Baacke:1993aj,Baacke:1994ix} 
for computing the fluctuation
determinants for the instanton in the Abelian Higgs model
in $1+1$ dimensions and for the sphaleron in the 
 SU(2) Higgs model in 3 dimensions, respectively.
When one tries to naively apply the Gel'fand-Yaglom method  to these cases
of topological solutions in gauge theories one encounters
a specific difficulty: the topological soliton modifies the 
centrifugal barriers and as a consequence the contribution
of the s-wave sector is infinite. The problem is avoided 
when using the Green's function method if the summation
over partial waves is carried out before the integration of the
Green's function. For the Gel'fand-Yaglom method
a modified approach is required; it is the aim of this
work to elucidate the problem and its solution.
At the same time we consider the relation between the methods,
analytically and numerically.
We use the Abelian Higgs model here mainly as a typical
example of a model with a topological soliton, the problem
is present for the sphaleron transition in the 
 electroweak SU(2) Higgs model as well.    

 The Abelian Higgs model in $(1+1)$ has found considerable
attention since on the one hand it shares certain features with the
electroweak theory,  and on the other hand it is simple enough
to serve as a theoretical and numerical laboratory.
In the context of the baryon number violation 
the high temperature sphaleron transition in this model has been studied
\cite{Grigoriev:1989je,Tang:1997cy} for which exact classical 
solutions and an exact expression of the sphaleron determinant 
\cite{Bochkarev:1987wg,Bochkarev:1989vu} are known, thus providing a complete
one-loop semiclassical
transition rate which can be studied numerically 
on the lattice, e.g. by
measuring the fluctuations of the Chern-Simons number.

Another prominent feature of the model is the existence
of instanton solutions \cite{Nielsen:1973cs,deVega:1976mi} 
which give rise
again to fluctuations in
the topological charge of the vacuum and thereby
to baryon number violation.  
In the dilute gas approximation for the instantons
transition rate, or equivalently the density of 
instantons in the Euclidean plane, is given by\cite{Coleman85}
\begin{equation}                                         \label{rate}
\Gamma =
 \frac{S(\phi_{cl})}{2\pi } 
{\cal D}^{-1/2} \exp (-S(\phi_{cl})-S_{ct}(\phi_{cl}))
\end{equation}
to one-loop accuracy. Here $S(\phi_{cl})$ is the
instanton action. The coefficient ${\cal D}$
represents the effect of quantum fluctuations around the instanton
configuration and arises from the Gaussian approximation to
the functional integral. This is the object
whose computation we will consider here.
It is given in general form by
\begin{equation}             \label{det}
{\cal D}
=\frac{\det'( \calm)  }{\det( \calm^0 )}
=\exp(2 S_{eff}^{1-loop}) \; ,
\end{equation}
the second equation relating it to the
one-loop effective action.
The operators $\calm$ are the fluctuation operators obtained
by taking the second functional derivative of the action
at the instanton and vacuum background field configurations.
The prime on the determinant implies omitting of the two translation
zero modes. The first prefactor $S(\phi_{cl})/2\pi$ takes
into account the integration of the
translation mode collective coordinates. 
Finally, the counterterm action $S_{ct}$ in the exponent will
absorb the ultraviolet divergences of ${\cal D}$.
One may also include
a corresponding determinant for fermions, which for massless
fermions is even known analytically \cite{Nielsen:1976hs,Nielsen:1977aw}.
For finite masses is has been computed recently
\cite{Burnier:2005he}.

This paper is organized as follows:
In the next section we present the basic
equations for  the Abelian Higgs model and its instanton.
The fluctuation operator and its partial wave reduction
is presented in section \ref{fluctuations}. Two methods for computing 
fluctuation integrals, one based on integrating Euclidean 
Green' s functions, as it was used in Ref. \cite{Baacke:1994bk}, 
and the Gel'fand-Yaglom method are introduced in section
\ref{twomethods} and compared. This includes the application 
for a single channel problem, as present here in the
Faddeev-Popov sector, and to a coupled-channel problem,
as present here for the gauge-Higgs sector.
In section \ref{specifix} we adress some specific problems:
in subsection \ref{swave} we discuss
the s-wave problem which arises do to the topological
nature of the background field and which constitutes the main
purpose of this manuscript, 
in subsection \ref{zeromode} the zero mode problem which
has well-known solutions for both the Green's function and
the Gel'fand-Yaglom method, and 
in subsection \ref{renorm} the renormalization, which here amounts to
as simple subtraction.
The numerical results, in particular a comparison of both
methods and the final results for the effective action
are presented in section \ref{numerix}.
A summary and conclusions are presented in section \ref{summary}.

%%%%%%%%%%%%%%%%%%%%%%%%%%%%%%%%%%%%%%%%%%%%%%%%%%%%%%%%%%%%%%%%%%%%%%%%%

\setcounter{equation}{0}
\section{The Abelian Higgs model and its instanton}
\label{abelianhiggs}
\par

The Abelian Higgs model in (1+1) dimensions is
defined by the Lagrange density (written in the Euclidean form
relevant here)
\begin{equation}
{\cal L}=\frac{1}{4}(F_{\mu\nu})^2
+\frac{1}{2}|D_\mu\phi|^2+\frac{\lambda}
{4}\left(|\phi|^2-v^2\right)^2\; .  
\end{equation}

Here
\begin{eqnarray*}
F_{\mu\nu}&=&\partial_\mu A_\nu-\partial_\nu A_\mu \\
D_\mu&=&\partial_\mu-igA_\mu \\ 
\end{eqnarray*}

The particle spectrum consists of Higgs bosons of mass
$m_H^2=2\lambda v^2$ and vector bosons of mass $m_W^2=g^2v^2$.
Usually the Higgs and gauge sector are coupled to a fermionic sector and
displays fermion number violation violated by instantons.
We here omit this aspect entirely.

The model has instanton solutions which change the topological
charge
\be
q=\frac{g}{2\pi}\int d^2 x \epsilon_{\mu\nu}F_{\mu\nu}
\pkt\ee
If the density of instantons is sufficiently small
they can be treated in the dilute gas approximation and be described
as separate objects with  topological charge
by $q=\pm 1$.

A structure which exhibits such a topological charge and satisfies
the Euclidean equations of motion is given by the
Nielsen-Olesen vortex \cite{Nielsen:1973cs}. The spherically symmetric
ansatz for this solution is given by
\begin{eqnarray}
A_\mu^{cl}(x)&=&\frac{\varepsilon_{\mu\nu}x_\nu}{gr^2}A(r)\kma \\
\phi^{cl}(x)&=&vf(r)e^{i\varphi(x)} \; .
\end{eqnarray}

In order to have a purely real Higgs field one performs a
gauge transformation
\begin{eqnarray} \label{gaugetr}
\phi &\to& e^{-i\varphi}\phi\kma \nonumber \\
 A_\mu&\to& A_\mu-\partial_\mu\varphi/g \nonumber \\
\end{eqnarray}
to obtain the instanton fields in the singular gauge
\begin{eqnarray}
A^{cl}_\mu (x)&=&\frac{\varepsilon_{\mu\nu}x_\nu}
{gr^2}(A(r)+1) \kma \\
\phi^{cl}(x)&=&vf(r) \; .
\end{eqnarray}
In this gauge the fields take to their vacuum values as
$r=|\bfx| \to \infty$.

With this ansatz the Euclidean action takes the form
\begin{eqnarray}
S_{cl}&=&\pi v^2
 \int^{\infty}_{0}\!\!\! dr \left(\frac{1}{rm_W^2}\left(
\frac{dA(r)}{dr}\right)^2\!\!+r\left(\frac{df(r)}{dr}\right)^2\!\!
+\frac{f^2(r)}{r}\left(A(r)+1\right)^2\!\! \right. \nonumber \\ 
&+&\left. \frac{rm^2_H}{4}\left(
f^2(r)-1\right)^2\!\right)
\pkt\end{eqnarray}

For the case $M_H=M_W$ an exact solution to the
variational equation is known \cite{deVega:1976mi},  for which
the classical action takes the value $S_{cl}=\pi v^2$.
We will consider here the general case, however, for which
the classical equations of motion
\begin{eqnarray}
\left(\frac{\partial^2}{\partial r^2}+
\frac{1}{r}\frac{\partial}{\partial r}
-\frac{\left(A(r)+1\right)^2}{r^2}-
\frac{m^2_H}{2}\left(f^2(r)-1\right)
\right)f(r)&=&0 \kma \\
\left(\frac{\partial^2}
{\partial r^2}-\frac{1}{r}
\frac{\partial}{\partial r}-m^2_W f^2(r)\right)
A(r)&=&m^2_W f^2(r) ~~~~
\end{eqnarray}
have to be solved numerically.
 
Imposing the boundary conditions on the profile functions
\begin{equation}  \label{rb}
\begin{array}{rcccccr}
A(r)&\stackrel{\scriptscriptstyle{r\to 0}}
{\longrightarrow}& const\cdot r^2                       4
&,&A(r)&\stackrel{\scriptscriptstyle{r\to\infty}}
{\longrightarrow}&-1 \\ 
f(r)&\stackrel{\scriptscriptstyle{r\to 0}}
{\longrightarrow}&const\cdot r 
&,&f(r)&\stackrel{\scriptscriptstyle{r\to\infty}}
{\longrightarrow}&1 
\end{array}
\end{equation}
the Chern-Simons number is $1$ and the action is finite.

Since we will consider fluctuations around these solutions later
on, a good numerical accuracy for the profile functions
$f(r)$ and $A(r)$ is required. We have found that the
method used previously by Bais and Primack \cite{Bais:1975bq}
in order to obtain precise profiles for the 't Hooft-Polyakov
monopole is very suitable also in this context. The method is outlined
in Appendix \ref{baisprimack}.

%%%%%%%%%%%%%%%%%%%%%%%%%%%%%%%%%%%%%%%%%%%%%%%%%%%%%%%%%%%%%%%%%%%%%%%%%

\setcounter{equation}{0}
\section{Fluctuation operator and mode functions}
\label{fluctuations}
The fluctuation operator is
defined in general form as
\begin{equation}
\calm =  \frac{\delta^2S}{\delta \psi^*_i (x) \delta \psi_j (x')}
|_{\psi_k=\psi_k^{cl}}\kma
\end{equation}
where $\psi_i$ denotes the fluctuating fields and
$\psi_i^{cl}$ the ``classical'' background field
configuration; here these
will be the instanton and the vacuum configurations. If the fields
are expanded around the background configuration as
$\psi_i = \psi_i^{cl} + \phi_i$
and if the Lagrange density is expanded accordingly, then the
fluctuation operator is related to the second order Lagrange density
via
\begin{equation}
{\cal L}^{II} =  \frac{1}{2}  \phi^*_i \calm_{ij}  \phi_j \; .
\end{equation}

In terms of the fluctuation operators
$\calm$ on the instanton  and $\calm^0$ on the
vacuum backgrounds, the effective action
is defined as
\begin{equation}
S_{eff} = \frac{1}{2} \ln \left\{ \frac{\det' \calm}
{\det \calm^0} \right\} \; .
\end{equation}
For our specific model we expand as
\begin{eqnarray}
A_\mu&=&A_\mu^{cl}+a_\mu  \kma\\ 
\phi&=&\phi^{cl}+\varphi \; .
\end{eqnarray}

In order to eliminate the gauge degrees of
freedom we introduce, as in Ref. \cite{Kripfganz:1989vm},
 the background gauge function
\begin{equation}
{\cal F}(A)=\partial_\mu A_\mu+\frac{ig}{2}\left((\phi^{ cl})
^\ast\phi-\phi^{cl}\phi^\ast\right) \kma
\end{equation}
which leads in the Feynman background gauge to the gauge-fixing
Lagrange density
\begin{eqnarray}
{\cal L}_{GF}^{I\hspace{-.05cm}I}&=
&\left(\frac{1}{2}{\cal F}^2(A)\right)
^{I\hspace{-.05cm}I} \nonumber \\ 
&=&\frac{1}{2}(\partial_\mu a_\mu)^2
-\frac{ig}{2}a_\mu(\varphi\partial_\mu\phi^{cl}
+\phi^{cl}\partial_\mu\varphi
-\varphi^\ast\partial_\mu\phi^{cl}-\phi^{cl}
\partial_\mu\varphi^\ast) \\
&&-\frac{g^2}{8}(\phi^{cl})^2(\varphi-\varphi^\ast)^2  \; .
\nonumber \end{eqnarray}
The associated Fadeev-Popov Lagrangian becomes
\begin{equation}
{\cal L}_{FP} =\frac{1}{2} \eta^\ast(-\partial^2+g^2
\left(\phi^{cl}\right)^2)\eta \; .
\end{equation}
In terms of the real components $\varphi= \varphi_1+i\varphi_2$
and $\eta = (\eta_1 + i \eta_2)/\sqrt{2}$ the second order Lagrange
density now becomes (omitting the superscript from $\phi^{cl}$ and
$A_\mu^{cl}$) 
\begin{eqnarray} \label{lag2}
\left({\cal L}+{\cal L}_{GF}+{\cal L}_{FP}\right)^
{I\hspace{-0.05cm}I}&=&
a_\mu\frac{1}{2}\left(-\partial^2+g^2\phi^2\right)
a_\mu \nonumber \\
&&+\varphi_1\frac{1}{2}\left(-\partial^2+g^2A_\mu^2+
\lambda\left(3\phi^2-
v^2\right)\right)\varphi_1  \nonumber \\  
&&+\,\varphi_2\frac{1}{2}\left(-\partial^2+g^2A_\mu^2+g^2\phi^2+
\lambda\left(\phi^2-v^2\right)\right)\varphi_2 \nonumber \\
&&+\,\varphi_2(gA_\mu\partial_\mu)\varphi_1
+\varphi_1(-gA_\mu\partial_\mu)\varphi_2  \\
&&+\,a_\mu(2g^2A_\mu\phi)\varphi_1
+\,a_\mu(2g\partial_\mu\phi)\varphi_2 \nonumber \\
&&+\eta_1\frac{1}{2}\left(-\partial^2+g^2\phi^2\right)\eta_1
+\eta_2\frac{1}{2}\left(-\partial^2+g^2\phi^2\right)\eta_2
\pkt \nonumber  
\end{eqnarray}
Specifying now the fluctuating fields
$ (\phi_1,\phi_2,\phi_3,\phi_4,\phi_5)$ as
$ (a_1,a_2,\varphi_1,\varphi_2,\eta_{12})$
the nonvanishing components of $\calm$ are

\bigskip
$\begin{array}{rcl@{\qquad}rcl}
\calm_{11}&=&\displaystyle -\partial^2 + g^2 \phi^2&
\calm_{22}&=&\displaystyle -\partial^2 + g^2 \phi^2 \\
\calm_{13}&=
& 2 g^2 A_1 \phi&\calm_{14}&=& 2 g \partial_1 \phi \\
\calm_{23}&=
&2 g^2 A_2 \phi&\calm_{24}&=& 2 g \partial_2 \phi \\
\calm_{33}&=&\displaystyle -\partial^2+g^2A_\mu^2 +
\lambda(3\phi^2-v^2)&
\calm_{34}&=&\displaystyle -g A_\mu \partial_\mu  \\
\calm_{44}&=&\displaystyle -\partial^2 +g^2 A_\mu^2 + g^2 \phi^2
+\lambda (\phi^2-v^2)& \calm_{43}&=& g A_\mu \partial_\mu \\
\calm_{55}&=&-\partial^2+g^2 \phi^2 \; .&
&&
\end{array}$ 
\bigskip

The Faddeev-Popov fluctuations, labelled with the subscript $5$,
represent an single channel system,
while the gauge-Higgs fluctuations, labelled with subscripts $1-4$,
form a $4\times 4$ coupled channel
system. It is understood that the contribution of the Faddeev-Popov
operator $\calm_{55}$ enters with a negative sign and a factor 
$2$ into the definition of the effective action.
The fluctuation operators for the instanton and vacuum
background are now obtained by substituting the corresponding
classical fields.
The vacuum fluctuation operator for the gauge-Higgs sector
 becomes a diagonal
matrix of Klein-Gordon operators with masses
$(M_W,M_W,M_W,M_H,M_W)$. It is convenient
to introduce a potential $\cal V$ via
\begin{equation}
\calm = \calm^0 + {\cal V} \; .
\end{equation}
The potential ${\cal V}$ will be specified below after partial
wave decomposition.

The fluctuation operator $\calm$ can be decomposed 
into partial waves and its determinant decomposes
accordingly,
\begin{equation}
\ln \det \calm = \sum_{n=-\infty}^{+\infty}
\ln \det {\bf M}_n
\pkt\end{equation}
We introduce the following partial wave decomposition for
fields 
\begin{eqnarray*}
\vec{a}&=&\sum_{n=-\infty}^{+\infty}
b_n(r)\left(\begin{array}{c}\cos\varphi \\\sin\varphi
\end{array}\right)\frac{e^{in\varphi}}{\sqrt{2\pi}}+ic_n(r)\left(
\begin{array}{c}-\sin\varphi\\\cos\varphi\end{array}\right)
\frac{e^{in\varphi}}{\sqrt{2\pi}}  \kma\\
\varphi_1&=&\sum_{n=-\infty}^{+\infty}
h_n(r)\frac{e^{in\varphi}}{\sqrt{2\pi}}\kma  \\
\varphi_2&=&\sum_{n=-\infty}^{+\infty}
\tilde{h}_n(r)\frac{e^{in\varphi}}{\sqrt{2\pi}}\kma  \\
\eta_{12}&=
&\sum_{n=-\infty}^{+\infty} g_n(r)\frac{e^{in\varphi}}{\sqrt{2\pi}}
\; .
\end{eqnarray*}

After inserting these expressions into the
Lagrange density and using the reality conditions for
the fields one finds that the following combinations are
real relative to each other and make the fluctuation operators
symmetric:
\begin{eqnarray*}
F^n_1(r)&=&\frac{1}{2}(b_n(r)+c_n(r)) \kma \\
F^n_2(r)&=&\frac{1}{2}(b_n(r)-c_n(r)) \kma \\
F^n_3(r)&=&\tilde{h}_n(r)  \kma\\
F^n_4(r)&=&ih_n(r)  \kma\\
F^n_{5}(r)&=&g_n(r)\pkt
\end{eqnarray*}
Writing the partial fluctuation operators - omitting the index
$n$ in the following - as
\begin{equation}
{\bf M} = {\bfM}^0 + {\bf V} \; ,
\end{equation}
the free operators ${\bf M}^0$ become diagonal matrices with
elements
\begin{equation}
M^0_{ii}= -\frac{d^2}{dr^2}-\frac{1}{r}\frac{d}{dr}
+\frac{n_i^2}{r^2} + M_i^2
\kma\end{equation}
where $(n_i) = (n-1,n+1,n,n,n)$ and $(M_i)=(M_W,M_W,M_W,M_H,M_W)$.

The potential $\bfV$ takes the elements

\bigskip
$\begin{array}{rcl@{\qquad}rcl}
\bfV_{11}^n&=&m_W^2\left(f^2-1\right) &\bfV_{12}^n&=&0\\
\bfV_{13}^n&=&\sqrt{2}m_W^{}f'&
\bfV_{14}^n&=&\displaystyle\sqrt{2}m_W^{}f
\frac{A+1}{r}\\
\bfV_{22}^n&=&\bfV_{11}^n&\bfV_{23}^n&=&\bfV_{13}^n\\
\bfV_{24}^n&=&-\bfV_{14}^n&
\bfV_{33}^n&=&\displaystyle
\frac{(A+1)^2}{r^2}+\frac{m_H^2}{2}\left(f^2-1\right)
+m_W^2\left(f^2-1\right)\\
\bfV_{34}^n&=&
\displaystyle -2\frac{A+1}{r^2}n&\bfV_{44}^n&=&
\displaystyle\frac{(A+1)^2}
{r^2}+\frac{3}{2}m_H^2\left(f^2-1\right)  \\
\bfV_{55}^n&=&m_W^2\left(f^2-1\right)&\bfV_{i5}&=&0 \; .\\
\end{array}$
\bigskip 

Choosing the dimensionless variable $M_Wr$ one realizes that
the fluctuation operator depends only on the ratio
$\xi=M_H/M_W$, up to an overall factor $M_W^2$ which cancels in
the ratio with the free operator.

We will need in the the Euclidean fluctuation modes
$f_{n,i}^{\alpha\pm}(r,\nu^2)$, which we will denote as ``mode
functions'' in the following.
They satisfy
\be
\left[\left(M^0_{ii}+\nu^2\right) \delta_{ij}+ \bfV^n_{ij}\right]
f^{\alpha\pm}_{n,j}=0
\pkt\ee
The superscript $\alpha$ labels  $4$ linearly independent solutions,
the subscript $i$ labels the $4$ components, $n$ refers to the partial
wave.
The superscript $+$ denotes a solution regular (i.e. exponentially
decreasing) as $r\to \infty$, the superscript $-$ denotes
a solution regular at $r=0$.
The corresponding
free ($\bfV=0$) solutions are the
modified Bessel functions $B^\pm_{n_i}(\kappa_i r)$ with
\bea
B^+_{n_i}(\kappa_i r)&=&K_{n_i}(\kappa_ir) \kma
\\
B^-_{n_i}(\kappa_i r)&=&I_{n_i}(\kappa_ir)
\kma\eea
where $\kappa_i=\sqrt{\nu^2+M_i^2}$.

It is convenient to rewrite the mode functions as
\be \label{hdef}
f^{\alpha\pm}_{n,i}(r,\nu^2)=B^\pm(\kappa_i r)\left[\delta^\alpha_i
+h^{\alpha\pm}_{n,i}(r,\nu^2)\right]
\pkt\ee
The functions $h^{\alpha\pm}_{n,i}(r,\nu^2)$ then satisfy
\bea \nonumber
&&\hspace{-10mm}\left\{\frac{d^2}{dr^2}+\left[\frac{1}{r}+2\kappa_i
\frac{B_{n_i}^{\pm'}(\kappa_i r)}{
B_{n_i}^\pm(\kappa_i r)}\right]\frac{d}{dr}\right\}h^{\alpha\pm}_{n,i}(r,\nu^2)
\\&&\hspace{10mm}=\bfV^n_{ij}(r)\frac{B^\pm_{n_j}(\kappa_jr)}{B^\pm_{n_i}(\kappa_ir)}
\left[\delta^\alpha_j + h^{\alpha\pm}_{n,j}(r,\nu^2)\right]   
\label{hequations}
\pkt\eea
We note that the functions $h^{\alpha\pm}_{n,i}$ collect terms
of first and higher order $\bfV^n$. The first order can be obtained by 
solving
\bea \nonumber
&&\hspace{-10mm}\left\{\frac{d^2}{dr^2}+\left[\frac{1}{r}+2\kappa_i
\frac{B_{n_i}^{\pm'}(\kappa_i r)}{
B_{n_i}^\pm(\kappa_i r)}\right]\frac{d}{dr}\right\}
h^{(1)\alpha\pm}_{n,i}(r,\nu^2)
\\&&\hspace{10mm}=\bfV^n_{ij}(r)\frac{B^\pm_{n_j}(\kappa_jr)}
{B^\pm_{n_i}(\kappa_ir)}
\delta^\alpha_j \label{h1equation}   
\pkt\eea

%%%%%%%%%%%%%%%%%%%%%%%%%%%%%%%%%%%%%%%%%%%%%%%%%%%%%%%%%%%%%%%%%%%%%%%%%

\setcounter{equation}{0}
\section{Two approaches to computing the logarithm of the
fluctuation determinant}
\label{twomethods}
\subsection{Method I: Integration of the Green's function}
\label{methodI}
We consider a partial differential 
 operator in two dimensions $\calm $ of the form
\be
\calm = -\Delta_2 + m^2+\calv(r)
\pkt\ee
In the Abelian Higgs model such an operator appears
in the Faddeev-Popov sector. The gauge-Higgs sector will
be considered later.
The potential $\calv$ has been assumed to be spherially  symmetric
($r=|\bfx|$), so that the 
space of eigenfunctions
is separable into partial waves of the form
\be
\psi_\gamma(\bfx)=R_\gamma(r)e^{in\varphi}
\ee
with
\be
\bfM_n R_\gamma(r)=\left[-\frac{d^2}{dr^2}-\frac{1}{r}\frac{d}{dr}
+\frac{n^2}{r^2}+\calv(r)\right]R_\gamma(r)=
-\omega_\gamma^2 R_\gamma(r)
\kma\ee
where we have introduced the partial wave reduction $\bfM_n$
of $\calm$ \footnote{
We use discrete notation for the eigenvalue spectrum, either this
can be realized by introducing a boundary at some large value $R$ of $r$
or it may simply be considered as a formal way of presentation.}. 
We further assume that $\calv(r)\to 0$ as $r\to \infty$ sufficiently
fast, e.g. to be of finite range, and is  nonsingular.
The index $\gamma=(j_n,n)$ is a multi-index, consisting of a radial
quantum number $j_n$ and an angular momentum (``magnetic'') quantum number $n$.
 We define the Euclidean Green's function of this operator via
\bea \nonumber
\calg(\bfx,\bfx',\nu^2)&=&\sum_\gamma
\frac{\psi_\gamma(\bfx)\psi_\gamma^\dagger(\bfx')}
{\omega_\gamma^2+  \nu^2}
=\sum_n e^{in(\varphi-\varphi')}
\sum_{j_n}\frac{R_{j_n}(r)R^\dagger_{j_n}(r')}
{\omega_\gamma^2+  \nu^2}
\\&=&\sum_n e^{in(\varphi-\varphi')}\bfG_n(r,r',\nu^2)
\eea
satisfying
\be 
(\calm_x+ \nu^2) \calg(\bfx,\bfx',\nu^2)=\delta^2(\bfx-\bfx')
\ee

The logarithm of the fluctuation determinant is defined as
\be
\calj(\nu^2)=\ln \frac{\det (\calm+ \bfI \nu^2)}{
\det(\calm_0+ \bfI \nu^2)}\kma
\ee
where $\bfI$ is the unit operator. Using the identity $\ln \det = \tr \ln $
the logarithm  can be calculated as 
\be
\calj(\nu^2)=
 \sum_\gamma \ln\frac{\omega_\gamma^2+  \nu^2}
{\left(\omega_{\gamma,0}\right)^2+  \nu^2} \kma
\ee
where $\omega_{\gamma,0}^2$ denotes the eigenvalues of the ``free'' operator
\be
\calm_0=-\Delta_2+m^2
\pkt\ee

In order to obtain $\calj(\nu^2)$ we begin by integrating the Green's function $\calg(\bfx,\bfx,\nu^2)$ over $\bfx$:
\be
\calg (\nu^2)\equiv\int d^2x\;\calg(\bfx,\bfx,\nu^2)=\sum_n \sum_{j_m}
\frac{1}{\omega_\gamma^2+  \nu^2}
\pkt\ee
Next we integrate with respect to $\nu^2$ from $0$ to $\Lambda^2$:
\bea
-\int_0^{\Lambda^2} d\nu^2  \calg(\nu^2)&=& 
-\int_0^{\Lambda^2} d\nu^2\sum_n \sum_{j_n}
\frac{1}{\omega_\gamma^2+\nu^2}\\
&=& 
\sum_n \sum_{j_n}\ln\frac{\omega_\gamma^2}{\omega_\gamma^2+\Lambda^2}
\pkt\eea
We now subtract the equivalent expression for the free operator
$\calm_0$ with the Green's function $\calg_0$ to obtain
\bea\nonumber
&&-\int_0^{\Lambda^2} d\nu^2 
\left[ \calg(\nu^2)-\calg_0(\nu^2)\right]
\\\nonumber 
&=&-\int_0^{\Lambda^2} d\nu^2\sum_m \sum_{n_m}
\left[\frac{1}{\omega_\gamma^2+\nu^2}-\frac{1}{(\omega_{\gamma,0})^2+\nu^2}
\right]
\\
&=& 
\sum_n \sum_{j_n}
\left[\ln\frac{\omega_\gamma^2}{\omega_{\gamma,0}^2}
-\ln\frac{\omega_\gamma^2+\Lambda^2}{\omega_{\gamma,0}^2+\Lambda^2}\right]
\\\nonumber
&=&\sum_n \bfJ_n(0,\Lambda^2)
\\&=&\nonumber\calj(0,\Lambda^2)
\pkt\eea
We now cannot longer avoid discussing the existence of the formal
expressions we have written down. Indeed already the expression
$\calg(\bfx,\bfx,\nu^2)$ does not exist.
An expression that {\em does}  exist is the integral over $r$ over the
partial wave Green's function:
\be
\bfG_n(\nu^2)\equiv \int dr r \bfG_n(r,r,\nu^2)
\pkt\ee
In order to obtain 
\be\label{seffGFunsub}
\calj(0,\Lambda^2)=-\int_0^{\Lambda^2}d\nu^2\sum_n 
\left[\bfG_n(\nu^2)-\bfG_{0,n}(\nu^2)\right]
\ee
 we may either sum
$\bfG_n(\nu^2)$ over $n$ and then integrate over $\nu^2$, or 
we may integrate over $\nu^2$ first and then do the sum over $n$. 
As long as $\Lambda^2$
is finite both ways lead to the same finite result.
But ultimately one wants to obtain $\calj(0,\infty)$. 
 The limit $\lim_{\Lambda^2\to \infty}
\calj(0,\Lambda^2)$ is naively expected to vanish, but 
in fact this is not the case,
rather one finds a logarithmic dependence on the cutoff $\Lambda^2$.
It reflects the fact that the  logarithm of the 
functional determinant is logarithmically
divergent. This can be found by computing the leading Feynman graphs
contributing to $\calj(\nu^2,\Lambda^2)$.

Within each partial wave one finds, for a 
nonsingular potential of finite range,
that $ \lim_{\Lambda^2\to \infty}
\bfJ_n(0,\Lambda^2)$ is finite. But  the sum
$\sum_n \bfJ_n(0,\infty)$ is logarithmically divergent, in contrast to the sum
$\sum_n(\bfJ_n(0,\Lambda^2))$. So the
operations of summation and taking the limit do not commute.  

The logarithmic divergence can be removed by subtracting the 
perturbative contribution of first order in $\calv$ from
$\bfJ_n(0,\Lambda^2)$. This will be discussed
in some detail in subsection \ref{renorm}. Then the summation over $n$ and
the integration over $\nu^2$ commute and the limit $\Lambda^2\to \infty$
can be taken.

We finally note that for practical computations one 
represents the partial wave Green's function
$\bfG_n(r,r',\nu^2)$ by Jost functions $f_n^\pm(r,\nu^2)$
which satisfy
\be
(\calm_n+\nu^2)f^\pm(r,\nu^2)=\left[-\frac{d^2}{dr^2}-\frac{1}{r}\frac{d}{dr}
  +\frac{n^2}{r^2}+\calv(r)+\nu^2\right]f^\pm_n(r,\nu^2)=0
\kma\ee
the boundary conditions
\be
f_n^-(r,\nu^2)\propto r^n\;\;\;{\rm for}\;\; r \to 0
\ee
and
\be
f_n^+(r,\nu^2)\to 0 \;\;\;{\rm for}\;\; r\to \infty
\pkt\ee
The solutions are normalized in such a way that the Wronskian is given by
\be
w(f^+,f^-)=f^+(r,\nu^2)\frac{d}{dr}f^-(r,\nu^2)-f^-(r,\nu^2)\frac{d}{dr}f^+(r,\nu^2)=\frac{1}{r^2}
\pkt \ee
These boundary conditions can be made more explicit
by writing
\bea
f^-_n(r,\nu^2)&=&I_n(\kappa r)\left[1+h^-_n(r,\nu^2)\right]
\\
f^+_n(r,\nu^2)&=&K_n(\kappa r)\left[1+h^+_n(r,\nu^2)\right]
\eea
with the condition
\be
\lim_{r\to \infty}h^\pm_n(r,\nu^2)=0
\pkt\ee
For $r\to 0$ the functions $h_n^\pm(r,\nu^2)$ go to  constants,
$h_n^\pm(r,\nu^2)\simeq h_n^\pm(0,\nu^2)+ O(r^2)$. 
The Wronskian condition entails
\be
1+h_n^+(0,\nu^2)=\frac{1}{1+h_n^-(0,\nu^2)}
\pkt\ee

With the mode functions $h^\pm_n$ defined in this way
 the partial wave Green's function is given by
\be
\bfG_n(r,r')=f_n^-(r_<)f_n^+(r_>)
\pkt\ee
Having computed the solutions $f^\pm$ for a particular set of
values of $n$ and $\nu^2$ one know's the Green' s function
for all values of $r$, so the $r$ integration can be done to obtain 
$\bfG_n(\nu^2)$, which is the basis for the subsequent steps. 

%%%%%%%%%%%%%%%%%%%%%%%%%%%%%%%%%%%%%%%%%%%%%%%%%%%%%%%%%%%%%%%%%%%%%%%%%

\subsection{Method II: the Gel'fand-Yaglom method}
\label{methodII}

The partial wave reduction of $\calm^2+\nu^2$
is an ordinary differential operator of second order,
and for computing the determinant of an ordinary differential
operator there is the Gel'fand-Yaglom theorem
stating that
\be  
\frac{\det{(\bfM_n+\nu^2\bfI)}}{\det{(\bfM_{0,n}+\nu^2\bfI)}}
=\lim_{r\to \infty}\frac{\tilde f^-_n(r,\nu^2)}
{\tilde f^-_{n,0}(r,\nu^2)}
\pkt
\ee
from which we then obtain 
\be
\bfJ_n(\nu^2)=\ln \lim_{r\to \infty}\frac{\tilde f^-_n(r,\nu^2)}
{\tilde f^-_{n,0}(r,\nu^2)}
\ee
Here the functions $\tilde f^-_n$ are identical to the functions
$f^-_n$ of the previous subsection {\em except} for the
normalization. The normalization is fixed by writing
\be
\tilde f^-_n(r,\nu^2)=I_n(\kappa r)\left[1+\tilde h^-_n(r,\nu^2)\right]
\ee
with the boundary condition
\be
\tilde h^-_n(0,\nu^2)=0
\pkt
\ee
We then have
\be
\bfJ_n(\nu^2)=\ln \left[1+\tilde h^-_n(\infty,\nu^2)\right]
\ee
In the Appendix we present heuristically two proofs of the
 theorem, one along the one given in Ref.
\cite{Coleman85}, and one which establishes a direct contact
with method I. The first version of the proof
is based on the condition for a bound
state  $\lim_{r\to \infty}f^-_n(r,\nu^2)=0$. 
Furthermore a  basic assumption
is that $\bfJ_n(\nu^2)\to 1$ as $\nu^2\to \infty$, i.e., that the
determinant of $\bfM_n$ tends towards the one of $\bfM_{n,0}$ 
in this limit, within each partial wave subspace. This is the
case for potentials of finite range. But then again we have
to sum over $n$ and this sum will be logarithmically
divergent. The renormalization for this case as for the
Green' s function method is discussed
in subsection \ref{renorm}.  

We have not introduced a cutoff here and stated the theorem
in its naive form. For the Faddeev-Popov sector this is indeed not necessary,
the perturbative subtractions can be done in the partial waves. Alternatively,
e.g. for comparing with the Green' s function method, 
one simply uses $\bfJ_n(0,\Lambda^2)=\bfJ_n(0)-\bfJ_n(\Lambda^2)$, where
the right hand side is evaluated by using the naive formula.

%%%%%%%%%%%%%%%%%%%%%%%%%%%%%%%%%%%%%%%%%%%%%%%%%%%%%%%%%%%%%%%%%%%%%%%%%

\subsection{Generalization to coupled channels}
\label{coupledchannels}
Both methods can be generalized to coupled channel systems 
with a spherical symmetry.
The operator $\calm$ and its partial wave reduction
$\bfM$ now become $N\times N$ matrices.
The solutions $f^\pm(,r\nu^2)$ are replaced by a fundamental
system of solutions $f^{\alpha\pm}_{n,i}(r,\nu^2)$ where
index $i$ labels the component and the index $\alpha$
labels the solution. Both indices run from $1$ to $N$, so that
the fundamental system can be considered to form an
$N\times N$ matrix labelled by $\alpha$ and $i$. 
We have already introduced these functions and
their differential equation at the end of section
\ref{fluctuations}.

If the Wronskian matrix within an angular momentum subspace
is given by
\be  
w^{\alpha\beta}=
\sum_i \left[f^{\alpha +}_i(r,\nu^2)\frac{d}{dr}f^{\beta -}_i
(r,\nu^2)-f^{\beta-}(r,\nu^2)\frac{d}{dr}f^{\alpha +}(r,\nu^2)\right]
=\frac{\omega^{\alpha\beta}}{r^2}
\kma \ee
then the Green's function is given by
\be\label{greenf}
\bfG_{n,ij}(r,r',\nu^2)={\omega^{-1}}_{\beta\alpha}
f^{\alpha +}_i(r,\nu^2)f^{\beta -}_j(r',\nu^2)
\pkt\ee 
From this Green' s function we are again able to compute the
fluctuation determinant using method I. We have
\be
\bfG_n(\nu^2)=\int r dr \bfG_{n,ii}(r,r,\nu^2) 
\ee
and the subsequent steps are performed as described above. 
In terms of the functions $h^{\alpha\pm}_{n,i}$,
see Eq. \ref{hdef} we have
\bea\nonumber
\bfG_{n,ii}(r,r,\nu^2)
&=&\left[h^{i+}_{n,i}(r,\nu^2)
+h^{i-}_{n,i}(r,\nu^2)\right.
\\&&\left.+\sum_\alpha h^{\alpha+}_{n,i}(r,\nu^2)
h^{\alpha-}_{n,i}(r,\nu^2)
\right]I_{n_i}(\kappa_ir)K_{n_i}(\kappa_ir)
\pkt \label{greenh}
\eea

As to the Gel'fand-Yaglom method we note that the condition for
a bound state of the coupled-channel system is given by
\be
\lim_{r\to \infty}\det f^{\alpha -}_{n,i}(r,\nu^2)=0
\ee
where the determinant refers to the matrix of the
fundamental system labelled by the indices $\alpha$ and $i$.
Using a proof analogous to the one given in
Ref. \cite{Coleman85} the theorem  takes the form
\be
\bfJ_n(\nu^2)=\ln \lim_{r\to \infty}
\frac{ {\rm det} \tilde f_{n,i}^{\alpha -}(r,\nu^2)}
{{\rm det} \tilde f_{0,n,i}^{\alpha -}(r,\nu^2)}
\ee
where the matrix of free solutions
$\tilde f_{0,n,i}^{\alpha -}$ can be taken as
${\rm diag}\{I_{n_i}(\kappa_i r)\}$ with $\kappa_i=\sqrt{\nu^+m_i^2}$.
Writing the solutions $\tilde f_{n,i}^{\alpha -}(r,\nu^2)$ in the form
\be
\tilde f_{n,i}^{\alpha -}(r,\nu^2)=\left(\delta^\alpha_i
+\tilde h_{n,i}^{\alpha -}(r,\nu^2)\right)I_{n_i}(\kappa_i r)
\ee
the boundary condition for the functions 
$\tilde h_{n,i}^{\alpha -}(r,\nu^2)$ is 
\be 
\lim_{r\to 0} \tilde h_{n,i}^{\alpha -}(r,\nu^2)=0
\ee
and the Gel'fand-Yaglom method yields
\be
\bfJ_n(\nu^2)=\ln\det\left\{\delta^\alpha_i
+\tilde h_{n,i}^{\alpha -}(\infty,\nu^2)\right\}
\pkt
\ee
This has been used, e.g., in \cite{Baacke:1995bw}, for computing the
fluctuation determinant for bubble nucleation in the $SU(2)$ Higgs
model at large temperature.

We have again stated the theorem in its naive form, without
cutoff. If the potential $\bfV^n$ is well-behaved, we again have
$\bfJ_n(0,\Lambda^2)=\bfJ_n(0)-\bfJ(\Lambda^2)$.
 The situation changes
if we consider the fluctuation determinant for a topological soliton,
indeed some matrix elements of the potential $\bfV_n$  in the
gauge-Higgs sector are singular as $r\to 0$. This will be discussed
in subsection \ref{swave}.

%%%%%%%%%%%%%%%%%%%%%%%%%%%%%%%%%%%%%%%%%%%%%%%%%%%%%%%%%%%%%%%%%%%%%%%%%

\setcounter{equation}{0}
\section{Specific problems}
\label{specifix}

\subsection{The case of topological solitons}
\label{swave}
If one analyzes the operator $\bfM_n$ for the case of
the Abelian instanton one finds that the centrifugal barriers
are modified in relation to the nontrivial
winding number. Indeed the potential $\bfV^n$ contains 
terms proportional to $1/r^2$ in the $33$, $44$ and $34$ and 
$43$ components.
While in the topologically trivial vacuum sector
the pattern of centrifugal barriers near $r=0$ for the gauge-Higgs system is 
$\{n_i\}=\{n-1,n+1,n,n\}$  within the subspace of angular momentum $n$,
 in the instanton sector this
becomes $\{\tilde n_i\}=\{n-1,n+1,n+1,n-1\}$. For $n=0$
we have $\{n_i\} =\{1,1,0,0\}$ which is distorted to
$\{\tilde n_i\}=\{1,1,1,1\}$.
As a consequence the partial waves near $r=0$ no longer behave
as the Bessel functions $I_{n_i}(\kappa r)$, and the functions
$h^{\alpha\pm}_{n,i}(r,\nu^2)$ no longer become constant
near $r=0$. For $n\neq 0$ two mixtures of the components 
with $i=3,4$ behave as $1/r$ and $r$, respectively; for $n=0$ both
Higgs components behave as $r$.

In Refs. \cite{Baacke:1994bk} and \cite{Baacke:1993aj,Baacke:1994ix} 
the method I
was used in order to compute the fluctuation determinant for the
Abelian instanton in $1+1$ and the of sphaleron in $3$ dimensions
(the high-temperature limit of the $3+1$ dimensional theory),
respectively. The summation over partial waves was done before
the integration over $\nu^2$ and the cutoff was sent to infinity
after suitable subtractions. However, if one does the $\nu^2$ 
integration first then one finds that the $n=0$ contribution becomes infinite
as $\Lambda^2\to \infty$, even after the suitable perturbative subtractions. 
For the Gel'fand-Yaglom method there is no integration over $\nu^2$,
one cannot interchange it with the summation over $n$, and one gets
the partial wave contributions at once. Indeed one finds that the $s$-wave 
contribution, if computed naively, is ill-defined.

The clue to the problem lies at first in the form in which the
theorem is applied. The naive expression
holds for ``well-behaved'' potentials.
If we consider the version of the theorem as derived 
in Appendix \ref{GYG} directly from the Green's function
method, it takes the form
\be
\bfJ_n(0,\Lambda^2)
=\ln \frac{\det {\left[\bfI+{\bf \tilde  h}^-_n(\infty,0)\right]}}
{\det{\left[\bfI+{\bf \tilde  h}^-_n(\infty,\Lambda^2)\right]}}
\ee
for a coupled channel problem. 

This form is suitable for all
partial waves except for the s-wave. We had required that at $r=0$
the determinant $\det {\left[\bfI+{\bf \tilde  h}^-_n(r,\nu^2)\right]}$
goes to unity. Relative to the free solutions, the Bessel functions,
the distorted centrifugal barriers lead, for $n\neq 0$,
to additional factors $r$ and $1/r$ for mixtures of the
Higgs field components $3$ and $4$. In the determinant these factors
compensate and so the normalization can be maintained. However, for $n=0$
where centrifugal barriers change from
$\{n_i\}_0 =\{1,1,0,0\}$ to
$\{\tilde n_i\}=\{1,1,1,1\}$ two additional factors $r$ appear 
in the Higgs sector, and the determinant
behaves as $r^2$ for $r\to 0$. So we have a problem
with the $s$-wave. In  the Green's function approach
the $s$-wave contribution is found to be finite as
long as $\Lambda^2$ remains finite. The solution of the puzzle
is to note that in a general normalization the theorem takes the form
\be
\bfJ_n(0,\Lambda^2)
=\ln \frac{\det {\left[\bfI+{\bf \hat  h}^-_n(\infty,0)\right]}
\det {\left[\bfI+{\bf \hat  h}^-_n(0,\Lambda^2)\right]}}
{\det{\left[\bfI+{\bf \hat  h}^-_n(\infty,\Lambda^2)\right]}
\det{\left[\bfI+{\bf \hat  h}^-_n(0,0)\right]}}
\pkt\ee 
Here we have set $r=0$ in the additional factors, but
in fact we should have taken the limit $r\to 0$ of their
ratio. Indeed, in the limit $r\to 0$
both $\det{\left[\bfI+{\bf \hat  h}^-_n(r,\Lambda^2)\right]}$
and $\det {\left[\bfI+{\bf \hat  h}^-_n(r,0)\right]}$
behave as $r^2$, the ratio does have a finite limit.
{\em So the introduction of the cutoff 
proves to be crucial:
it allows for the cancellation of the factor $r^2$.}
For suitable initial conditions at $r=0$ for the fundamental
system $h^{\alpha -}_n(r,\nu^2)$ the ratio can be made equal to 
unity independent of $\Lambda^2$.
By this modification of the numerical procedure the $s$-wave
contribution can be computed, for finite $\Lambda^2$,
using the Gel'fand-Yaglom method.
Once we have introduced the cutoff in the $s$ wave we are forced
to work with the cutoff in all partial waves, because we can let it
tend to infinity only after summation over all partial
waves. 

Even though now we have at least defined
the $s$-wave contribution there remains the fact that 
even after subtraction of the
perturbative first order contribution the $s$-wave
contribution gets infinite as $\Lambda^2\to \infty$. This
 singular behaviour is cancelled by the sum over the
higher partial waves. But this only works if the
summation over $n$ is performed before taking the
limit $\Lambda^2\to\infty$. This will be manifest in the
numerical results and will be displayed in section \ref{numerix}.

%%%%%%%%%%%%%%%%%%%%%%%%%%%%%%%%%%%%%%%%%%%%%%%%%%%%%%%%%%%%%%%%%%%%%%%%%

\subsection{Zero modes}
\label{zeromode}
The $m=1$ partial wave has a bound state at $\nu^2=0$.
As the partial wave is degenerate with $m=-1$ the bound state has
a  twofold degeneracy. These two zero modes correspond to
the translational collective degrees of freedom. The logarithm
of their eigenvalues would be infinite and has to be removed.

For the Gel'fand-Yaglom method two methods have been proposed, by
V. Kiselev and J.B. \cite{Baacke:1993ne} and by Dunne and Min
\cite{Dunne:2005rt}. The second one is semianalytic but
has not yet been adapted to
a coupled-channel problem, so we will use the first one:
Due to the zero mode the ratio of determinants
\be
\frac{\det( {\bfM_1 + \bfI \nu^2)}  }{\det( \bfM_{0,1} + \bfI \nu^2)}
\ee
has a zero at $\nu^2=0$. The factor $\nu^2$ can be removed by taking
the derivative of this expression at $\nu^2=0$. This
derivative can be computed numerically by computing the determinant
at two sufficiently small values of $\nu^2$.   
So we have
\be
\bfJ_1 (0)=\ln \frac{\det' {\bfM_1}  }{\det{ \bfM_{0,1}}}
=\ln \frac{d}{d\nu^2}\left[\frac{\det' {{\bfM_1}+ \bfI \nu^2}  }{\det{ \bfM_{0,1}
+ \bfI \nu^2}}\right]_{\nu^2=0}
\ee
For the $n=1$ partial wave, as for all higher partial waves,
$\bfJ_n(0,\Lambda^2)$ is computed simply as the difference between
$\bfJ_n(0)$ and $\bfJ_n(\Lambda^2)$, evaluated with the naive formula.
For the $n=1$ partial wave it is worth mentioning that
the procedure for removing the zero mode is applied to $\bfJ(0)$ only.
In taking the derivative we introduce a scale, and indeed the
procedure of removing the mode gives a dimension to the
effective action, this is discussed below. 

In the Green's function method one directly computes the
logarithm of the determinant and we have to use another procedure 
for removing the zero mode.  
Extending the integral
$-\int d\nu^2 \bfG_1(\nu^2)$ from
$\nu^2=\epsilon^2$ to $\nu^2=\Lambda^2$ the zero mode
manifests itself by divergence $\ln \epsilon^2$ as $\epsilon^2\to 0$.
It is this divergence which has to be removed. A straightforward
idea is to simply subtract the pole and to  compute
\be
-\int_{\epsilon^2}^{\Lambda^2} \left(\bfG_1(\nu^2)- \frac{1}{\nu^2}
-\bfG_{0,1}(\nu^2)\right)
+\ln \Lambda^2 \pkt
\ee
Numerically the subtraction as we have just defined it is not suitable.
The integral over $\bfG_1(\nu^2)-\bfG_{0,1}(\nu^2)$ 
is convergent as
$\Lambda^2\to \infty$, but not so the one over the pole term which
we have subtracted. A rather simple solution 
of this problem consists in 
subtracting $1/\nu^2$ only in the interval $[0,1]$. Then the
integral we have subtracted just produces the $\ln \epsilon^2$
which is to be removed and $\ln \Lambda^2$ is replaced by
$\ln 1 =0$. 

We note that this procedure introduces the logarithm
of a dimensionful quantity, while before the
argument of the logarithm was dimensionless.
Here all numerical computations are performed setting $m_W^2=1$.
So, as the exponential of the $-1/2 \log \det$ appears in the
transition rate, and as the translation mode is twofold degenerate,
the rate will be in units $M_W^2$. 

%%%%%%%%%%%%%%%%%%%%%%%%%%%%%%%%%%%%%%%%%%%%%%%%%%%%%%%%%%%%%%%%%%%%%%%%%

\subsection{Renormalization}
\label{renorm}
As we have already mentioned the limit $\Lambda^2 \to \infty$ of
$\calj(0,\Lambda^2)$ does not exist. This is due to the
divergences of quantum field theory. 
In the present case of a two-dimensional
model the divergences are just tadpole diagrams proportional
to $\phi^2$ and $\phi$ which have to be subtracted. We describe
the procedure for the gauge-Higgs sector, the application to
the Faddeev-Popov sector is obvious.

In the Green' s function method the expression for
$\calj(0,\Lambda^2)$, Eq. (\ref{seffGFunsub}) 
is replaced by
\be\label{seffGFsub}
\calj(0,\Lambda^2)_{sub}
=-\int_0^{\Lambda^2}d\nu^2\sum_n 
\left[\bfG_n(\nu^2)-\bfG_{0,n}(\nu^2)
-\bfG^{(1)}_n(\nu^2)\right]
\ee
where of course
\be 
\bfG_n^{(1)}=\int dr r \bfG_{n,ii}(r,r,\nu^2)
\pkt\ee
The diagonal components of the first order Green's function 
are given by
\be
G_{n,ii}(r,r)=\left[h^{(1)i+}_{n,i}(r,\nu^2)+
h^{(1)i-}_{n,i}(r,\nu^2)\right]I_{n_i}(\kappa_ir)K_{n_i}(\kappa_ir)
\ee
and the first order part of the functions $h^{i\pm}_{n,i}$ is
obtained by solving Eq. (\ref{h1equation})which for the diagonal 
elements reduces to
\be
\hspace{-10mm}\left\{\frac{d^2}{dr^2}+\left[\frac{1}{r}+2\kappa_i
\frac{B_{n_i}^{\pm'}(\kappa_i r)}{
B_{n_i}^\pm(\kappa_i r)}\right]\frac{d}{dr}\right\}
h^{(1)i\pm}_{n,i}(r,\nu^2)
=\bfV^n_{ii}(r)
\label{h1diagequation}   
\pkt\ee
The boundary condition is $h^{(1)i\pm}_{n,i}(\infty,\nu^2)\to 0$.
There is one important point, however: there cannot be divergences
with external gauge legs, i.e. proportional to
$a^\mu A_\mu$, as in a gauge theory the is no mass
counter term for the gauge field. 
Indeed these tadpole contributions 
are compensated by second order terms, in scalar QED in $4$
dimensions this corresponds the cancellation of quadratic divergences
proportional to $A^\mu A_\mu$. {\em 
So when computing  $h^{(1)i\pm}_{n,i}$ 
using Eq. (\ref{h1diagequation}) the terms $(A+1)^2/r^2$
have to be omitted  in $\bfV^n_{33}$ and
$\bfV^n_{44}$}, all divergences are proportional to
$f^2(r)-1$.

For the Gel'fand-Yaglom method the procedure consists again in
removing the first order part from
\be
\bfJ_n(0,\Lambda^2)
=\ln \frac{\det {\left[\bfI+{\bf \tilde  h}^-_n(\infty,0)\right]}}
{\det{\left[\bfI+{\bf \tilde  h}^-_n(\infty,\Lambda^2)\right]}}
\ee
The functions  
$\tilde h^{(1)i\pm}_{n,i}(r,\nu^2)$ are again  solutions
of Eq. (\ref{h1diagequation}), but with the boundary condition
$\tilde h^{(1)i\pm}_{n,i}(r,\nu^2)\to 0$ as $r\to 0$.  
Again the gauge field terms have to be omitted in the potential.
The subtracted expression is simply
\bea\nonumber
\left[\bfJ_n(0,\Lambda^2)\right]_{sub}
&=&\ln \frac{\det {\left[\bfI+{\bf \tilde  h}^-_n(\infty,0)\right]}}
{\det{\left[\bfI+{\bf \tilde  h}^-_n(\infty,\Lambda^2)\right]}}
\\&&- \sum_i\left[
\tilde h^{(1)i\pm}_{n,i}(\infty,0)-
\tilde h^{(1)i\pm}_{n,i}(\infty,\Lambda^2)\right]
\pkt\eea

%%%%%%%%%%%%%%%%%%%%%%%%%%%%%%%%%%%%%%%%%%%%%%%%%%%%%%%%%%%%%%%%%%%%%%%%%

\setcounter{equation}{0}
\section{Numerical results}
\label{numerix}
The classical profiles were computed using the Bais-Primack method
which is described in Appendix \ref{baisprimack}.
We have used 2000 grid point for $x=m_W r$  in the interval
$[0,30]$, the grid was not equidistant, but the 
interval length was chosen to increase by a
factor $1.005$ between two neighbouring intervals, so as to
have small intervalls at small $r$ and larger ones in the
asymptotic region.

\begin{figure}
\begin{center}
\includegraphics[scale=0.4]{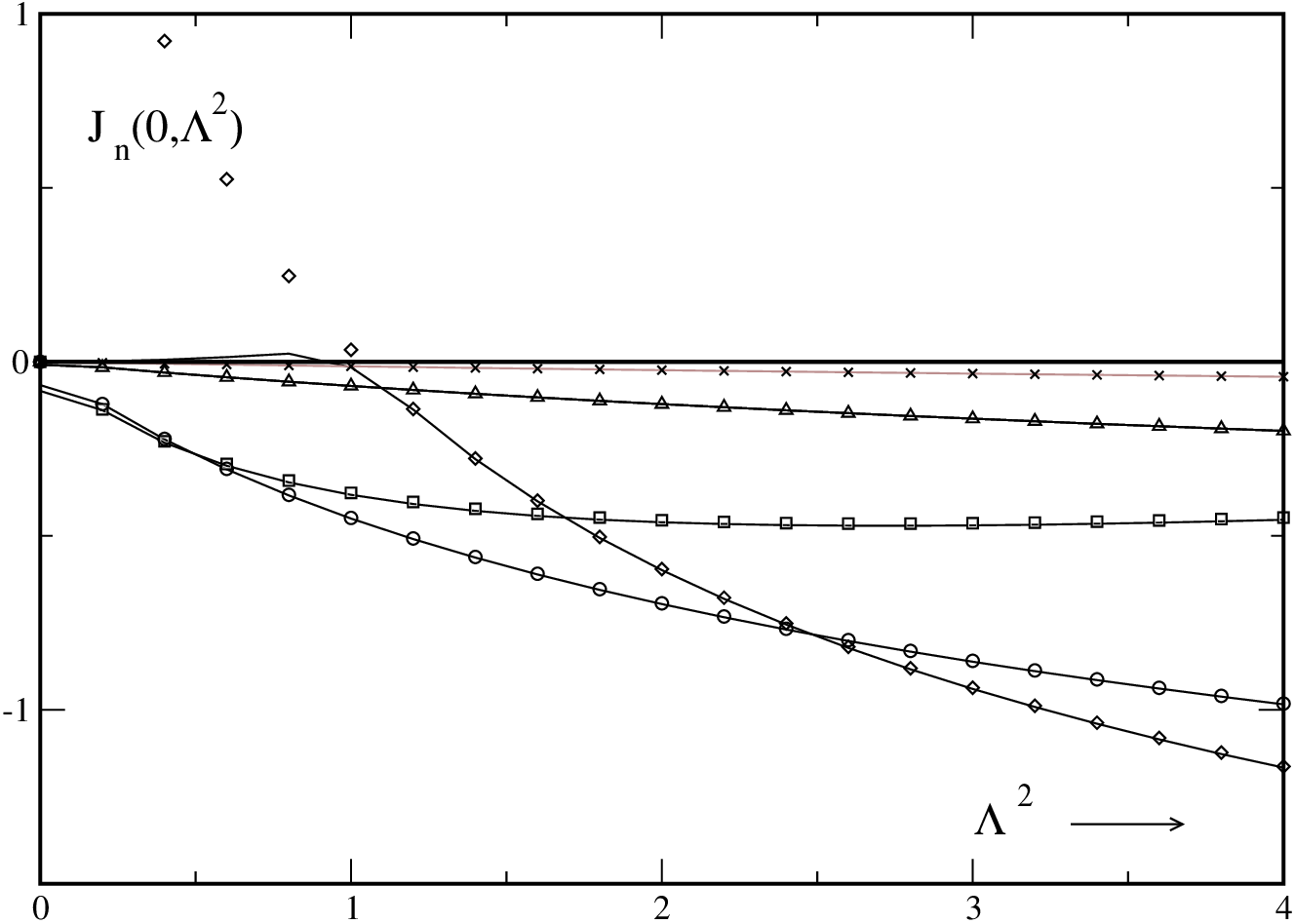}
\vspace*{3mm}
\end{center}
\caption{\label{comparesmalllambda}
Dependence of the partial wave functional determinant
on the cutoff: small $\Lambda$; we display the quantity $J_n(0,\Lambda^2)$
as a function of $\Lambda^2$. The solid lines are the results of the
Green' s function approach, the symbols are those of the Gel'fand-Yaglom
method; squares: $n=0$; diamonds: $n=1$, circles: $n=2$; triangles: $n=5$,
crosses: $n=10$.} 
\end{figure}

The methods of computing the fluctuation determinants has
been described in the previous sections, this discussion already
incorporates the numerical procedure.
We have compared the two methods for computing functional
determinants analytically. Of course this should reflect itself
in the numerical computations. The quantity to
be computed and compared is $J_n(0,\Lambda^2)$ and its partial
wave sum $\calj(0,\Lambda^2)$. In the following we 
just display results for $n \ge 0$, of course those for $-n$
are identical to those for $n$.

In Figure \ref{comparesmalllambda} we display the contributions of
various partial waves $J_n(0,\Lambda^2)$ for small $\Lambda^2$.
The results of the Green's function method and of the Gel'fand-Yaglom
method agree within drawing accuracy, and in fact to better than
$1\%$ for all values. A difference is found for the $n=1$ partial
wave, due to our translation mode subtraction; the result are expected
to agree for $\Lambda^2 > 1$, and they do. The singularity at small
$\Lambda^2$ found in the Gel'fand-Yaglom results (diamonds) is due to 
the fact that the translation mode is removed from $J_n(0)$ but not 
from $J_n(\Lambda^2)$; this is correct.

\begin{figure}
\begin{center}
\includegraphics[scale=0.4]{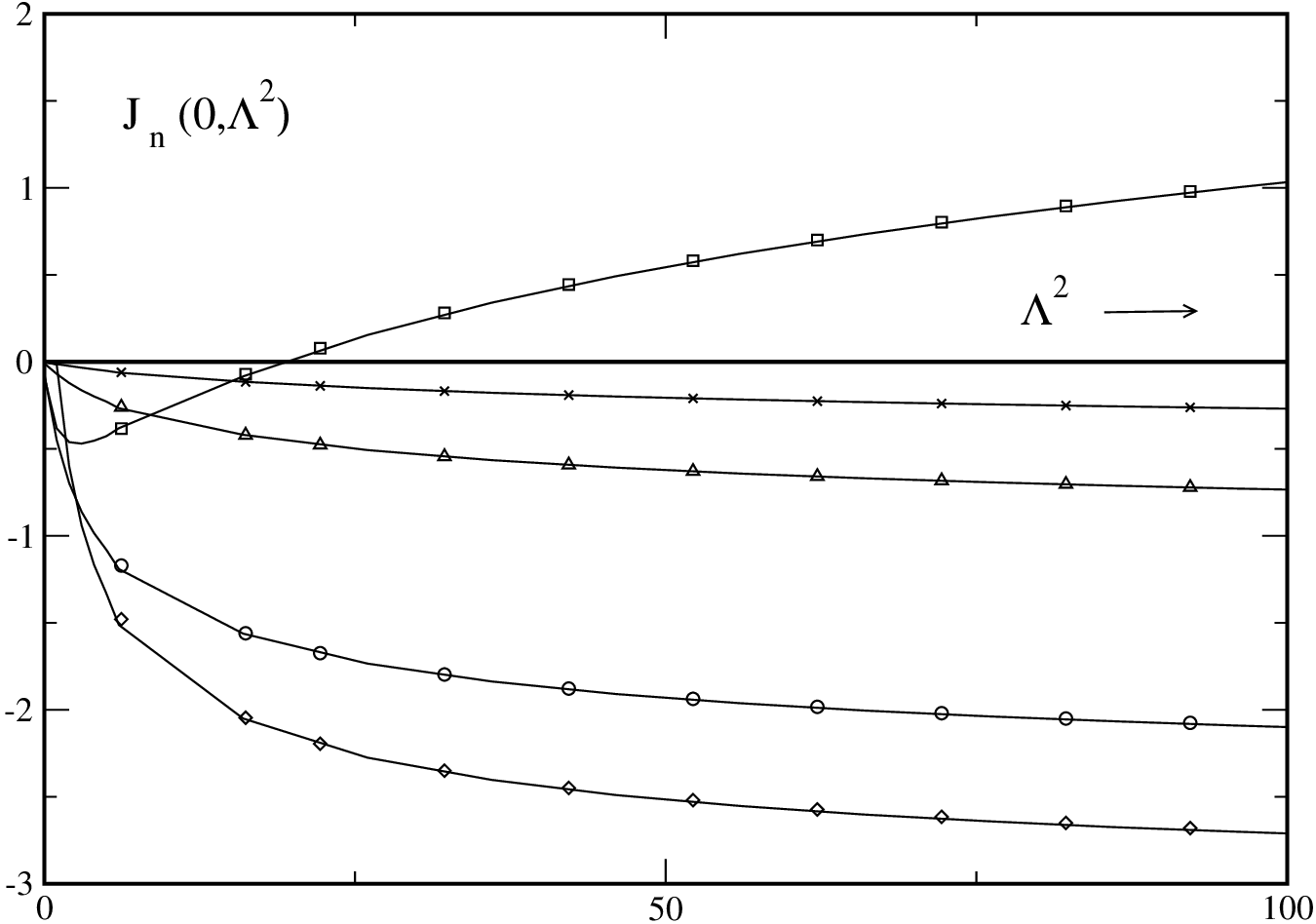}
\vspace*{3mm}
\end{center}
\caption{\label{comparelargelambda}
Dependence of the partial wave functional determinant
on the cutoff: large $\Lambda$; we display the quantity $J_n(0,\Lambda^2)$
as a function of $\Lambda^2$. The solid lines are the results of the
Green' s function approach, the symbols are those of the Gel'fand-Yaglom
method; squares: $n=0$; diamonds: $n=1$, circles: $n=2$; triangles: $n=5$,
crosses: $n=10$.}
\end{figure}

In Figure \ref{comparelargelambda} we again display the quantity
$J_n(0,\Lambda^2)$, this time for large $\Lambda$. In these
results, as in the ones of the previous figure the first order 
perturbative contribution is not yet subtracted. So neither of the 
contributions is expected to have a finite limit as $\Lambda^2\to \infty$,
they should behave as $\ln \Lambda^2$. One sees that the $s$ wave, $n=0$ 
contribution changes sign and evolves in the positive direction,
opposite to the other ones.

\begin{figure}
\begin{center}
\includegraphics[scale=0.4]{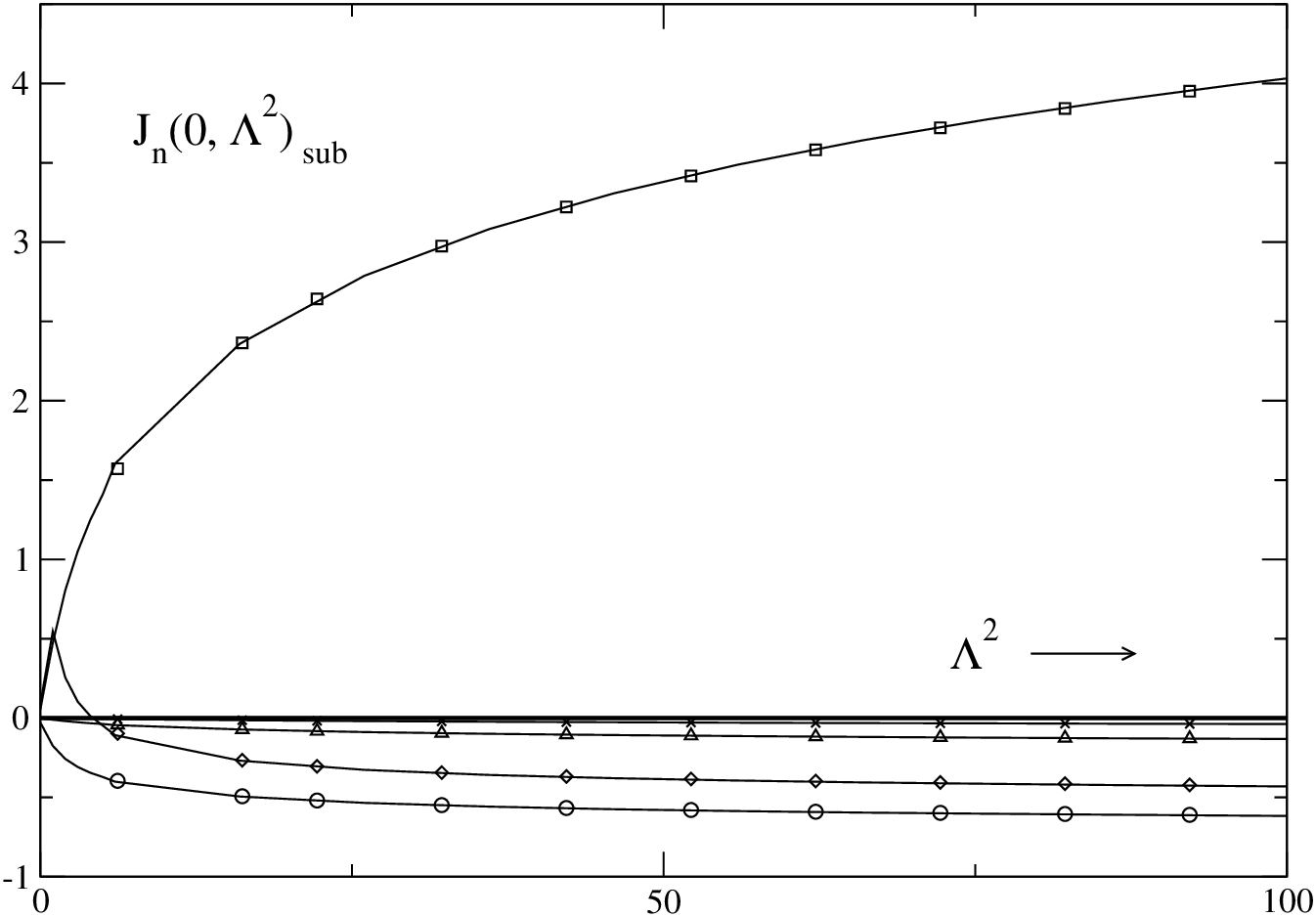}
\vspace*{3mm}
\end{center}
\caption{\label{comparelargelambdasub}
Dependence of the partial wave functional determinant
on the cutoff: large $\Lambda$; we display the quantity
 $J_n(0,\Lambda^2)$
as a function of $\Lambda^2$ after subtraction
of the first perturbative order. The lines and symbols refer to the same
partial waves in the previous figure.}
\end{figure}

If one subtracts the first order perturbative contribution the picture changes,
as displayed in Figure \ref{comparelargelambdasub}. Now all partial waves
with $n\neq 0$ have finite limits as $\Lambda^2\to \infty$, but not so the
one with $n=0$. This is the manifestation of the $s$ wave problem.
At finite $\Lambda^2$ the sum over partial waves is convergent, and 
in the limit $\Lambda^2\to \infty$ the
singular behaviour of the $s$-wave is compensated by the other
partial waves.

\begin{figure}
\begin{center}
\includegraphics[scale=0.4]{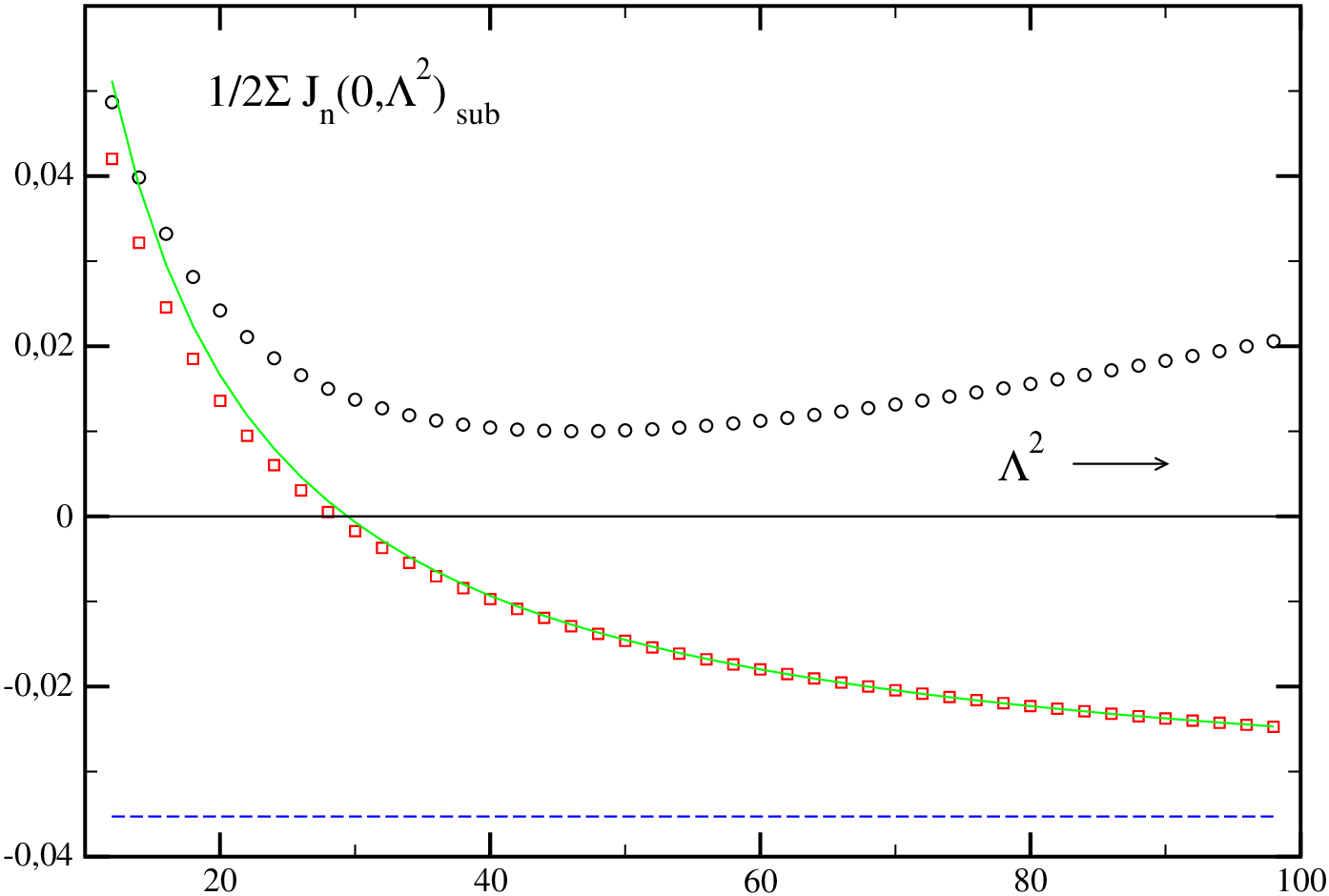}
\vspace*{3mm}
\end{center}
\caption{\label{asfitsplot}
Cutoff dependence of partial wave sums:
the  sums $1/2\Sigma_n J_n(0,\Lambda^2)_{sub}$
as a function of $\Lambda^2$; circles: summation up to $\bar n=35$,
squares: extrapolated sums; solid line: the fit
$-.353+1.037/\Lambda^2$, dashed line: asymptotic value $-.353$.
The sums include the Faddeev-Popov contributions.} 
\end{figure}

Having computed the quantities $J_n(0,\Lambda^2)_{sub}$ we have to
do the sum over partial waves and then let $\Lambda^2\to \infty$.
We find that the terms $J_n(0,\Lambda^2)_{sub}$ behave
as $n^{-3}$ for the gauge-Higgs system and as $n^{-5}$
for the Faddeev-Popov system. If the terms have been computed
up to some $\bar n$, we extrapolate by fitting the terms
$\bar n -5$ to $\bar n$ to $A n^{-3}+B n^{-4}+C n^{-5}$
and $A n^{-5}+B n^{-6}+C n^{-6}$, respectively. Then we append
the sum from $\bar n +1$ to $\infty$ using this fit. This procedure
 has been used in previous publications, it has been
checked here by varying $\bar m$ between
$20$ and $35$ to give reliable results.
The sums up to $n=35$ and the extrapolated sums are plotted
in Fig. \ref{asfitsplot} for $10 < \Lambda^2 < 100$.
Here the Faddeev-Popov contributions are included.
Obviously neither of these are independent of $\Lambda^2$.
The sum with the fixed upper limit $\bar n=35$ first decreases and
then starts to increase. This is a consequence of the
cancellation of the $s$-wave divergence by the other partial
waves. As $\Lambda^2$ increases, more and more partial waves are
necessary for this compensation, so with a fixed number of partial
waves this cannot work. The extrapolated sum is not constant either,
but it can be fitted to a behaviour $S_\infty+c/\Lambda^2$. Subasymptotic
corrections of order $1/\Lambda^2$ are expected.
We consider the number $S_\infty$ as the asymptotic value, to
be identified with the effective action. It is obvious that
the cancellation between the $s$-wave and the higher partial waves
becomes more an more delicate if $\Lambda^2$ increases, so it is
not suitable to choose even higher values of
$\Lambda^2$ to get a better estimate for $S_\infty$.

\begin{figure}
\begin{center}
\includegraphics[scale=0.4]{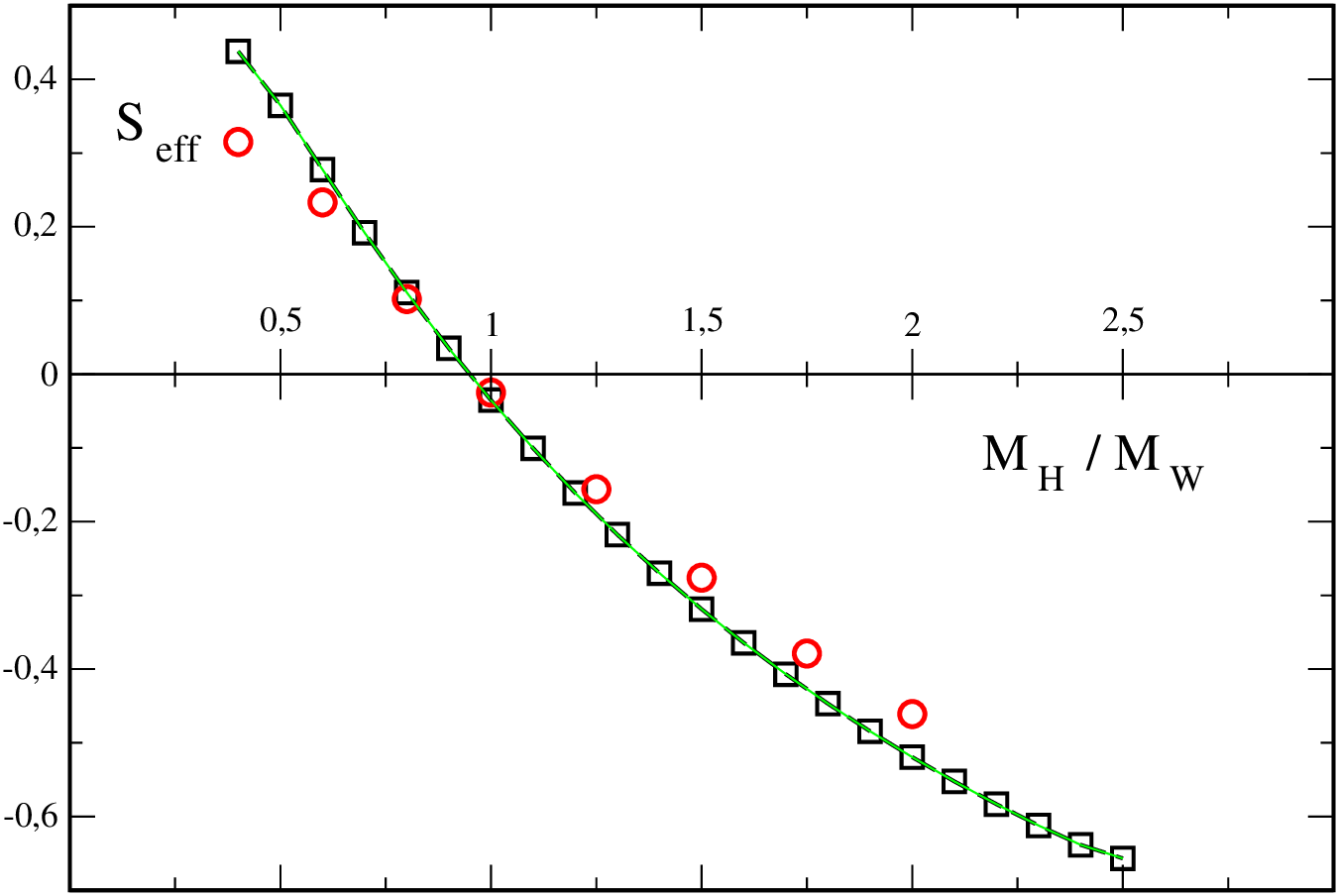}
\vspace*{3mm}
\end{center}
\caption{\label{seff}
The effective action; squares: our results; circles: the results
of Ref. \cite{Baacke:1994bk}.}
\end{figure}

The results for the effective action  
\be
S_ {eff}=\lim_{\Lambda^2\to\infty}
\frac{1}{2}J_n(0,\Lambda^2)_{sub}
\ee
found by the procedure we have described above
are displayed in Fig. \ref{seff}. We also plot the results obtained in our
previous publication with Torsten Daiber \cite{Baacke:1994bk}.
These results are consistent with the present ones within the error
of $0.07$ estimated in Ref. \cite{Baacke:1994bk}, except for the
value at $\xi=.4$. We do not intend to give
a detailed error estimate here. From varying the maximal value of 
the angular momentum and the cutoffs we think that
the error is around $0.01$, i.e. within ``drawing accuracy''. 
The Green's function and Gel'fand-Yaglom methods
produce consistent results within an error margin of better than
$1\%$. However we still have to rely on extrapolations, and this produces
some uncertainty, which may be systematic.   
 
An interesting problem appears when $\xi=M_H/M_W > 2$.  The
nondiagonal terms $V_{34}=V_{43}$ behave as $\exp(-M_W r)$. So
if $f^{\alpha-}_3(r,\nu^2)\simeq exp(-\kappa_W r)$ it contributes with
a behaviour $\exp(-(\kappa_W +M_W)r$ to $f^{\alpha-}_4(r,\nu^2)$.
Now $f^{\alpha-}_4$ is supposed to behave as
$\exp(-\kappa_H r)$. The cross term dominates over this behaviour for
$\nu^2<M_H^2(M_H^2-4M_W^2)/4M_W^2$. This defines a real interval
for $\nu$ if $M_H>2M_W$. It found indeed that for $M_H>2M_W$ and small
$\nu^2$ some of the functions $h^{\alpha-}_i$ increase exponentially
with the expected behaviour. One finds (numerically) that
these contributions cancel in the determinant, both methods
still produce consistent results up to $\xi=M_H/M_W\simeq 2.5$. However, this
cancellation is delicate numerically and for larger values $\xi$
the numerical procedure breaks down, the results become
inconsistent. So one would have to find a suitable 
modification of the numerical procedure in order to maintain
numerical reliability, here we limit ourselves to the range
$\xi < 2.5$.

%*******************************************************************

\setcounter{equation}{0}
\section{Summary}
\par\label{summary}
In this work we have addressed a problem that arises
in the fluctuation operator and in functional determinant
for external field configurations with nontrivial winding number.
A modification of the centrifugal barrier factors $n_i^2/r^2$
by the singular external field configuration necessitates
 modifications in the computation of functional determinants.
While these are relatively trivial when the computation
is carried out using the Green's function method, for the
Gel'fand-Yaglom they are less obvious. We have described
here both approaches. Indeed the handling of the
problem when using the Green's function method has led us
to the solution of the problem for the Gel'fand-Yaglom
method. It consists in introducing a cutoff, which before
was unnecessary after suitable perturbative subtraction.

We have presented the numerical comparison of both methods,
the results are found to agree  within an accuracy 
of better than $1\%$. The results for the one-loop
effective action agree with the previous calculation in
Ref. \cite{Baacke:1994bk} within the errors given there.

%%%%%%%%%%%%%%%%%%%%%%%%%%%%%%%%%%%%%%%%%%%%%%%%%%%%%%%%%%%%%%%%%%%%%%%%%

\begin{appendix}

\section{The Bais-Primack method}
\label{baisprimack}
\setcounter{equation}{0}
We shortly address the problem of finding reliable
classical solutions for the vortex background field.
As we have already mentioned, we use a method developed by
Bais and Primack \cite{Bais:1975bq}, which has to be adapted to the
vortex system.

We introduce two functions $F_A(x)$ and $F_f(x)$ with
$x=M_W r$ via
\bea
A(r)&=&-1+xF_A(x)
\\
f(r)&=&1+F_f(x)
\pkt\eea
The boundary condition for these new functions
are that  $f_A(x)\to 0$ and $F_f(x)\to 0$ as
$x\to\infty$. For $x\to 0$ they have to behave as
 $f_A(x)=1/x+O(x)$ and $F_f(x)=-1/x+O(x)$. 
They satisfy the differential equations
\bea
F_A''+\frac{1}{x}F_A'-\frac{1}{x^2}F_A-F_A&=&F_A(F_f^2-2F_f)
\\
F_f''+\frac{1}{x}F_f' -\xi^2 F_f &=&F_A^2(1+F_f)+\frac{\xi^2}{2}
F_f^2(3+F_f)
\pkt\eea
which have been written in such a way that
the differential operators on the left hand side
\bea
\cald_A&=&\frac{d^2}{dx^2}+\frac{1}{x}\frac{d}{dx}-\frac{1}{x^2}-1
\\
\cald_f&=&\frac{d^2}{dx^2}+\frac{1}{x}\frac{d}{dx}-\xi^2
\eea
are of the Bessel type. Their Green' s functions
are given by
\bea
G_A(x,x')&=&I_1(x_<)K_1(x_>)
\\
G_f(x,x')&=&I_0(\xi x_<)K_0( \xi x_>)
\eea
and we have the integral equations
\bea
F_A(x)&=&F_A^0(x)-\int_0^\infty x dx  G_A(x,x')\left[
F_A(x')(F_f^2(x')+2F_f(x'))\right. 
\\\nonumber &&\left.-\cald_A F_A^0(x')\right]
\\
F_f(x)&=&F_f^0(x)-\int_0^\infty x dx  G_f(x,x')\left[
F_A^2(x')(1+F_f(x')\right.
\\\nonumber &&\left.+\frac{\xi^2}{2}F_f^2(x')\left(3+F_f(x')\right)
-\cald_f F_f^0(x')\right]
\pkt\eea
The functions $F_f^0$ und $F_A^0$ have been introduced in order to
provide the solutions with the right boundary conditions.
They have to satisfy the same boundary conditions as the
solutions we are looking for.
We have chosen
\bea
F_A^0(x)&=&K_1(x)\frac{1+a_1 x}{1+a_2 x}
\\
F_f^0(x)&=&-K_1(\xi x)\frac{\xi x}{1+f_1 x}
  \pkt\eea
with suitable parameters $a_1,a_2$ and $f_1$ the iteration
of the integral equations produces solutions with an
accuracy of $10^{-9}$ after around $150$ iterations.

%%%%%%%%%%%%%%%%%%%%%%%%%%%%%%%%%%%%%%%%%%%%%%%%%%%%%%%%%%%%%%%%%%%%%%%%%

\section{Coleman's proof of the Gel'fand-Yaglom theorem}
\setcounter{equation}{0}
\label{colemansproof} 
In the book by Coleman \cite{Coleman85}
the Gel'fand-Yaglom theorem is stated in the following way:
Let $f(r,\nu^2)$ and $ f^0(r,\nu^2)$
denote the solutions of
\be
(\bfM+\nu^2) f(r,\nu^2)=0  
\ee
and
\be
(\bfM^0 +\nu^2)f^0 (r,\nu^2)=0  \; ,
\ee
respectively, on the interval $[0,\infty]$, 
with regular boundary conditions at $r=0$.
Let these solutions be normalized such that
\be
\lim_{r \to 0} \frac{f(r,\nu^2)}{f^0(r,\nu^2)} = 1 \;.
\ee
Then the following equality holds:
\be \label{flucdef}
\frac{\det (\bfM +\nu^2)}{\det (\bfM^0+\nu^2)}
= \lim_{r \to \infty} \frac { f(r,\nu^2)}{f^0(r,\nu^2)} 
 \pkt\ee
The argument consists of two parts:
\\
(i) as the bound state condition for the functions
$f(r,\nu^2)$ and $f^0(r,\nu^2)$ is given by
$f(\infty,-\omega_\alpha^2)=0$ and 
$f^0(\infty,-{\omega_\alpha^{0}}^2)=0$, and as
the determinants can be written, in the basis of
eigenstates, as products with factors $\nu^2+\omega_\alpha^2$
and similarly for the determinant of $\bfM^0+\nu^2$
the right hand and left hand sides are meromorphic functions
with identical poles and zeros. 
\\
(ii) therefore, if furthermore both sides become unity as $\nu^2\to \infty$,
they are identical. This condition holds for a large class
of potentials, in particular nonsingular potentials
of finite range. Intuitively one expects that for 
large $\nu^2$ the potential then becomes irrelevant and that the
solutions $f(r,\nu^2)$ and $f^0(r,\nu^2)$ become identical, 
the condition my be checked by perturbative expansion.

Generalized to a coupled $(n \times n)$ system the theorem can be
stated in the following way:

Let ${\bf f}(r,\nu^2)$ and ${\bf f}^0(r,\nu^2)$
denote the $(n \times n)$ matrices formed by 
$n$ linearly independent solutions 
$f_i^\alpha(r,\nu^2)$ and $f_i^{\alpha 0}(r,\nu^2)$ of
\be
(\bfM_{ij}+\nu^2) f_j^\alpha (r,\nu^2) =0  
\ee
and
\be
( \bfM_{ij}^0 +\nu^2) 
f^{\alpha 0}_j (r,\nu^2) =0  \; ,
\ee
respectively, with regular boundary conditions at $r=0$.
The lower index denotes the $n$ components, the
different solutions are labelled by the Greek upper index.
Let these solutions be normalized such that
\be
\lim_{r \to 0} {\bf f}(r,\nu^2)({\bf f}^0(r,\nu^2))^{-1} = {\bf 1} \;.
\ee
Then the following equality holds:
\be \label{coupledflucdef}
\frac{\det (\bfM +\nu^2)}{\det (\bfM^0+\nu^2)}
= \lim_{r \to \infty} \frac {\det {\bf f}(r,\nu^2)}{\det
{\bf f}^0(r,\nu^2)} 
\ee
where the determinants on the left hand side are determinants
in functional space, those on the right
hand side are ordinary determinants of the $n \times n$ matrices
defined above.

The argument goes as before, one just has to replace the bound
state condition for the one-channel problem by the condition
\be
\lim_{r\to\infty}\det {\bf f}(r,\nu^2)=0
\ee
as suited for a coupled-channel problem.

%%%%%%%%%%%%%%%%%%%%%%%%%%%%%%%%%%%%%%%%%%%%%%%%%%%%%%%%%%%%%%%%%%%%%%%%%

\section{Gel'fand-Yaglom and the Green's function}
\label{GYG}\setcounter{equation}{0}
In section  \ref{twomethods} we have
have introduced two numerical methods for computing the functional 
determinant of an operator $\calm$, or rather of operators
$\bfM_n$, the reduction of the operator $\calm$ to a subspace
of definite angular momentum. We will here connect the two methods
directly, without going back to the eigenfunctions of this operator
which are not used in either of the methods. The connection
between the methods has been discussed in Ref. \cite{Kleinert:1998rz},
we adapt their aproach  to the radial and coupled-channel operators 
which we have consider here.

We go back to the diffential equation satisfied by the mode functions
$f^\pm(r,\kappa)$
\be
\left[-\frac{1}{r}\frac{d}{dr}r\frac{d}{dr}
+\frac{n^2}{r^2}+V(r)+\nu^2\right]f_n^\pm{r,\nu^2}=0
\ee
Taking the derivative with respect to $\nu^2$ of the differental
equation for $f^-_n$ we have
\be
\left[-\frac{1}{r}\frac{d}{dr}r\frac{d}{dr}
+\frac{n^2}{r^2}+V(r)+\nu^2\right]\frac{d}{d\nu^2}f_n^-(r,\nu^2)
+f_n^-(r,\nu^2)=0
\ee
Multiplying with $f^+(r,\nu^2)$ and using the differential equation for 
$f_n^+$ we obtain
\bea \nonumber
&&\left[\frac{1}{r}\frac{d}{dr}r\frac{d}{dr}f^+(r,\nu^2)\right]
\frac{d}{d\nu^2}f_n^-(r,\nu^2)
-f^+(r,\nu^2)\frac{d}{d\nu^2}\frac{1}{r}\frac{d}{dr}r\frac{d}{dr}f^-(r,\nu^2)
\\
&&=-f_n^+(r,\nu^2)f_n^-(r,\nu^2)=-\bfG_n(r,r,\nu^2)
\eea
Integrating the Green's function over $r$ we have
\bea \nonumber
&&-\int_0^\infty r dr \bfG_n(r,r,\nu^2)=
\\\nonumber
&&=-\int_0^\infty r dr \frac{1}{r}\frac{d}{dr}
\left\{\left[r\frac{d}{dr}f_n^+(r,\nu^2)\right]
\frac{d}{d\nu^2}f_n^-(r,\nu^2)
-f_n^+(r,\nu^2)\frac{d}{d\nu^2}r\frac{d}{dr}f_n^-(r,\nu^2)\right\}
\\
&&=\left\{\left[r\frac{d}{dr}f_n^+(r,\nu^2)\right]
\frac{d}{d\nu^2}f_n^-(r,\nu^2)
-f_n^+(r,\nu^2)\frac{d}{d\nu^2}r\frac{d}{dr}f_n^-(r,\nu^2)\right\}\biggl|_0^\infty   \label{master}
\eea
In fact we have to compute the integral over $\bfG_n-\bfG_{0,n}$ and with the 
boundary conditions and normalization we have introduced the contributions
of the two Green's functions cancel each other at the upper integration
limit. Near the lower integration limit the functions $f^\pm_n(r,\nu^2)$
behave as $(1+h^+_n(r,\nu^2))K_n(\kappa r)$ and 
 $(1+h^-_n(r,\nu^2))I_n(\kappa r)$ with $\lim_{r\to 0}
(1+h_n^+(r,\nu^2))(1+h_n^-(r,\nu^2))=1$. Therefore the parts where
the derivatives $d/d\nu^2$ act on the Bessel functions cancel with
the free contribution and we remaining term is given by
\bea \nonumber
&&-(1+h^+_n(0,\nu^2))\frac{d}{d\nu^2}(1+h^-_n(0,\nu^2)
w(K_n(\kappa r),I_n(\kappa r))
\\
&&
=
\frac{d h^-_n(0,\nu^2)/d\nu^2}{1+h^-_n(0,\nu^2)}
=\frac{d}{d\nu^2}\ln\left[1+h^-_n(0,\nu^2)\right] 
\kma
\eea
where we have used
\be 
w(K_n(z)I_n(z))\equiv z \left[I_n(z)\frac{d}{dz}K_n(z)-
K_n(z)\frac{d}{dz}I_n(z)\right]=-1
\pkt\ee
In subsection \ref{methodII} we have introduced the functions
$\tilde f^-_n(r,\nu^2)$ which differ from the $f^-_n$ in the normalization.
Writing
\be
\tilde f^-_n(r,\nu^2)=I_n(\kappa r)\left[1+\tilde h^-_n(r,\nu^2)\right]
\ee
we have
\be
1+h_n^-(0,\nu^2)=\frac{1}{1+\tilde h_n^-(\infty,\nu^2)}
\ee
So we obtain
\bea
\bfJ_n(\nu^2,\Lambda^2)&=&-\int_{\nu^2}^{\Lambda^2} d\nu^{'2}
\left[\bfG_n(\nu^{'2})-\bfG_{0,n}(\nu^{'2})\right]
\\
&=&\int_{\nu^2}^{\Lambda^2}d\nu^{'2}\frac{d}{d\nu^{'2}}
\ln\left[1+h^-_n(0,\nu^{'2})\right] 
\\
&=&\ln\left[1+h^-_n(0,\nu^{'2})\right]\biggl|_{\nu^2}^{\Lambda^2}
\\
&=&-\ln\left[1+{\tilde h}^-_n(\infty,\nu^{'2})\right]\biggl|_{\nu^2}
^{\Lambda^2}
\\
&=&\ln \frac{1+\tilde h^-_n(\infty,\nu^2)}{1+\tilde h^-_n(\infty,\Lambda^2)}
\eea

Generalizing to coupled channels, choosing the Wronskian derterminant
$\omega^{\alpha\beta}=\delta^{\alpha\beta}$ we have
\bea 
&&-\int_0^\infty r dr \bfG_{n,ii}(r,r,\nu^2)=
\\\nonumber
&&=\left\{\left[r\frac{d}{dr}f_{n,i}^{\alpha +}(r,\nu^2)\right]
\frac{d}{d\nu^2}f_{n,i}^{\alpha -}(r,\nu^2)
-f_{n,i}^{\alpha +}(r,\nu^2)\frac{d}{d\nu^2}r\frac{d}{dr}f_n^{\alpha-}
(r,\nu^2)\right\}\biggl|_0^\infty   \label{mastercoupled}
\eea
The subsequent reasoning about the contributions of the upper
and lower integration limit is analogous.  We now have                  
\bea \nonumber
&&-(1+h^{\alpha+}_{ni}(0,\nu^2))\frac{d}{d\nu^2}
(1+h^{\alpha -}_{ni}(0,\nu^2)
w(K_{n_i}(\kappa r),I_{n_i}(\kappa r))
\\
&&=
\left[\bfI+{\bf h}^-_n(0,\nu^2)\right]^{-1}
\frac{d {\bf h}^-_n(0,\nu^2)}{d\nu^2}
=\frac{d}{d\nu^2}\ln\det \left[\bfI+{\bf h}^-_n(0,\nu^2)\right] 
\pkt
\eea
and by analogous steps as for the single-channel case we arrive at
\bea \nonumber
\bfJ_n(0,\Lambda^2)&=&-\int_0^{\Lambda^2} d\nu^2
\left[\bfG_{n,ii}(\nu^2)-\bfG_{0,n,ii}(\nu^2)\right]
\\\nonumber
&=&\int_0^{\Lambda^2}d\nu^2\frac{d}{d\nu^2}
\ln\det \left[\bfI+{\bf h}^-_n(0,\nu^2)\right] 
\\
&=&\ln \frac{\det {\left[\bfI+{\bf \tilde  h}^-_n(0,0)\right]}}
{\det{\left[\bfI+{\bf \tilde  h}^-_n(0,\Lambda^2)\right]}}
\eea
As we will see it is not always possible to normalize the fundamental
system at $r=0$. A more general expression, which treats the 
boundaries $r=0$ and $r\to \infty$ in a symmetrical way is given
by
\be
\bfJ_n(0,\Lambda^2)
=\ln \frac{\det {\left[\bfI+{\bf \hat  h}^-_n(\infty,0)\right]}
\det {\left[\bfI+{\bf \hat  h}^-_n(0,\Lambda^2)\right]}}
{\det{\left[\bfI+{\bf \hat  h}^-_n(\infty,\Lambda^2)\right]}
\det{\left[\bfI+{\bf \hat  h}^-_n(0,0)\right]}}
\kma
\ee
where now the matrix ${\bf \hat h}=\{\hat h^\alpha_n\}$ refers to a fundamental system
with an arbitrary normalization.

\end{appendix}
 
\bibliography{abinstdet}
\bibliographystyle{h-physrev4}

\end{document}